\documentclass[preprint2,apjl,numberedappendix,twocolappendix,appendixfloats]{emulateapj}

\usepackage[caption=false]{subfig}
\usepackage{amsmath}
\usepackage{footnote}
\bibpunct{(}{)}{;}{a}{}{,} 
\newcommand{\dif}{\mathrm{d}}

\def\nar{{New~A~Rev.}}          
\def\pasa{{PASA}}               

\shorttitle{Observations of Wide-field Effects in EoR Power Spectra}
\shortauthors{Thyagarajan et~al.}

\def\ASU{\altaffilmark{1}}
\def\ASUtxt{\altaffiltext{1}{Arizona State University, School of Earth and Space Exploration, Tempe, AZ 85287, USA}}

\def\myemail{\altaffilmark{*}}
\def\myemailtxt{\altaffiltext{*}{e-mail: t\_nithyanandan@asu.edu}}

\def\UW{\altaffilmark{2}}
\def\UWtxt{\altaffiltext{2}{University of Washington, Department of Physics, Seattle, WA 98195, USA}}

\def\SKASA{\altaffilmark{3}}
\def\SKASAtxt{\altaffiltext{3}{Square Kilometre Array South Africa (SKA SA), Park Road, Pinelands 7405, South Africa}}

\def\RU{\altaffilmark{4}}
\def\RUtxt{\altaffiltext{4}{Department of Physics and Electronics, Rhodes University, Grahamstown 6140, South Africa}}

\def\CfA{\altaffilmark{5}}
\def\CfAtxt{\altaffiltext{5}{Harvard-Smithsonian Center for Astrophysics, Cambridge, MA 02138, USA}}

\def\ANU{\altaffilmark{6}}
\def\ANUtxt{\altaffiltext{6}{Australian National University, Research School of Astronomy and Astrophysics, Canberra, ACT 2611, Australia}}

\def\CAASTRO{\altaffilmark{7}}
\def\CAASTROtxt{\altaffiltext{7}{ARC Centre of Excellence for All-sky Astrophysics (CAASTRO)}}

\def\Haystack{\altaffilmark{8}}
\def\Haystacktxt{\altaffiltext{8}{MIT Haystack Observatory, Westford, MA 01886, USA}}

\def\RRI{\altaffilmark{9}}
\def\RRItxt{\altaffiltext{9}{Raman Research Institute, Bangalore 560080, India}}

\def\MIT{\altaffilmark{10}}
\def\MITtxt{\altaffiltext{10}{MIT Kavli Institute for Astrophysics and Space Research, Cambridge, MA 02139, USA}}

\def\Curtin{\altaffilmark{11}}
\def\Curtintxt{\altaffiltext{11}{International Centre for Radio Astronomy Research, Curtin University, Perth, WA 6845, Australia}}

\def\Victoria{\altaffilmark{12}}
\def\Victoriatxt{\altaffiltext{12}{Victoria University of Wellington, School of Chemical \& Physical Sciences, Wellington 6140, New Zealand}}

\def\UWisc{\altaffilmark{13}}
\def\UWisctxt{\altaffiltext{13}{University of Wisconsin--Milwaukee, Department of Physics, Milwaukee, WI 53201, USA}}

\def\UMelbourne{\altaffilmark{14}}
\def\UMelbournetxt{\altaffiltext{14}{The University of Melbourne, School of Physics, Parkville, VIC 3010, Australia}}

\def\USydney{\altaffilmark{15}}
\def\USydneytxt{\altaffiltext{15}{The University of Sydney, Sydney Institute for Astronomy, School of Physics, NSW 2006, Australia}}

\def\CASS{\altaffilmark{16}}
\def\CASStxt{\altaffiltext{16}{CSIRO Astronomy and Space Science (CASS), PO Box 76, Epping, NSW 1710, Australia}}

\def\Tata{\altaffilmark{17}}
\def\Tatatxt{\altaffiltext{17}{National Centre for Radio Astrophysics, Tata Institute for Fundamental Research, Pune 411007, India}}

\begin{document}

\title{Confirmation of Wide-Field Signatures in Redshifted 21~cm Power Spectra}


\author{
Nithyanandan~Thyagarajan\ASU\myemail,
Daniel~C.~Jacobs\ASU,
Judd~D.~Bowman\ASU,
N.~Barry\UW,
A.~P.~Beardsley\UW,
G.~Bernardi\SKASA$^,$\RU$^,$\CfA,
F.~Briggs\ANU$^,$\CAASTRO,
R.~J.~Cappallo\Haystack, 
P.~Carroll\UW,
A.~A.~Deshpande\RRI, 
A.~de~Oliveira-Costa\MIT,
Joshua~S.~Dillon\MIT,
A.~Ewall-Wice\MIT,
L.~Feng\MIT,
L.~J.~Greenhill\CfA,
B.~J.~Hazelton\UW,
L.~Hernquist\CfA, 
J.~N.~Hewitt\MIT,
N.~Hurley-Walker\Curtin,
M.~Johnston-Hollitt\Victoria,
D.~L.~Kaplan\UWisc, 
Han-Seek~Kim\UMelbourne$^,$\CAASTRO,
P.~Kittiwisit\ASU,
E.~Lenc\USydney$^,$\CAASTRO,
J.~Line\UMelbourne$^,$\CAASTRO,
A.~Loeb\CfA,
C.~J.~Lonsdale\Haystack, 
B.~McKinley\UMelbourne$^,$\CAASTRO,
S.~R.~McWhirter\Haystack,
D.~A.~Mitchell\CASS$^,$\CAASTRO, 
M.~F.~Morales\UW, 
E.~Morgan\MIT, 
A.~R.~Neben\MIT,
D.~Oberoi\Tata, 
A.~R.~Offringa\ANU$^,$\CAASTRO, 
S.~M.~Ord\Curtin$^,$\CAASTRO,
Sourabh~Paul\RRI,
B.~Pindor\UMelbourne$^,$\CAASTRO,
J.~C.~Pober\UW,
T.~Prabu\RRI, 
P.~Procopio\UMelbourne$^,$\CAASTRO,
J.~Riding\UMelbourne$^,$\CAASTRO,
N.~Udaya~Shankar\RRI, 
Shiv~K.~Sethi\RRI,
K.~S.~Srivani\RRI, 
R.~Subrahmanyan\RRI$^,$\CAASTRO, 
I.~S.~Sullivan\UW,
M.~Tegmark\MIT,
S.~J.~Tingay\Curtin$^,$\CAASTRO, 
C.~M.~Trott\Curtin$^,$\CAASTRO,
R.~B.~Wayth\Curtin$^,$\CAASTRO, 
R.~L.~Webster\UMelbourne$^,$\CAASTRO, 
A.~Williams\Curtin, 
C.~L.~Williams\MIT,
J.~S.~B.~Wyithe\UMelbourne$^,$\CAASTRO
}

\ASUtxt
\UWtxt
\SKASAtxt
\RUtxt
\CfAtxt
\ANUtxt
\CAASTROtxt
\Haystacktxt
\RRItxt
\MITtxt
\Curtintxt
\Victoriatxt
\UWisctxt
\UMelbournetxt
\USydneytxt
\CASStxt
\Tatatxt
\myemailtxt


\begin{abstract}

We confirm our recent prediction of the ``pitchfork'' foreground signature in power spectra of high-redshift 21~cm measurements where the interferometer is sensitive to large-scale structure on all baselines. This is due to the inherent response of a wide-field instrument and is characterized by enhanced power from foreground emission in Fourier modes adjacent to those considered to be the most sensitive to the cosmological H~{\sc i} signal. In our recent paper, many signatures from the simulation that predicted this feature were validated against Murchison Widefield Array (MWA) data, but this key {\it pitchfork} signature was close to the noise level. In this paper, we improve the data sensitivity through the coherent averaging of 12 independent snapshots with identical instrument settings and provide the first confirmation of the prediction with a signal-to-noise ratio $>10$. This wide-field effect can be mitigated by careful antenna designs that suppress sensitivity near the horizon. Simple models for antenna apertures that have been proposed for future instruments such as the Hydrogen Epoch of Reionization Array and the Square Kilometre Array indicate they should suppress foreground leakage from the {\it pitchfork} by $\sim 40$~dB relative to the MWA and significantly increase the likelihood of cosmological signal detection in these critical Fourier modes in the three-dimensional power spectrum. 

\end{abstract}
 

\section{Introduction}\label{sec:intro}

The epoch of reionization (EoR) commenced following the formation of the first stars and galaxies. It is characterized by a period of non-linear growth of matter density perturbations and astrophysical evolution in the universe's history. The detection of redshifted 21~cm radiation of H~{\sc i} from this epoch is one of the most promising probes of the evolution of large-scale structure during this epoch \citep{sun72,sco90,mad97,toz00,ili02}.

Sensitive instruments such as the Square Kilometre Array (SKA), which will be capable of providing direct imaging of redshifted H~{\sc i}, are yet to become operational. In the meantime, the Hydrogen Epoch of Reionization Array\footnote{\url{http://reionization.org/}} (HERA), currently under development, will be much more advanced in its capability to detect and place definitive constraints on the reionization epoch relative to current instruments such as the Murchison Widefield Array \citep[MWA;][]{lon09,bow13,tin13}, the Low Frequency Array \citep[LOFAR;][]{van13}, and the Precision Array for Probing the Epoch of Reionization \citep[PAPER;][]{par10}, which have only enough sensitivity for a statistical detection of the signal \citep{bow06,par12a,bea13,dil13,thy13,pob14}.

The primary challenge to the detection of cosmological H~{\sc i} from the EoR arises from continuum emission from Galactic and extragalactic foreground objects, which is $\sim 10^4$ stronger than the desired signal. However, the inherent differences in spatial isotropy and spectral smoothness can be exploited to extract the cosmological signal from foreground contamination \citep[see, e.g.,][]{dim02,dim04,fur04,mor04,zal04,san05,fur06,mcq06,mor06,wan06,gle08}. Thus, a detailed characterization of foreground emission has become essential \citep{ali08,ber09,ber10,bow09,liu09,liu14a,liu14b,dat10,liu11,gho12,mor12,par12b,tro12,dil13,dil14,pob13,thy13,thy15}.

Our recent study \citep[][hereafter referred to as Paper~I]{thy15} used instrument and foreground models, for the first time with full sky coverage, in order to simulate actual EoR experiments more accurately than previous studies. Surprisingly, we found that foreground emission outside the primary beam field of view caused the most significant contamination of the Fourier modes considered the most sensitive for detecting the cosmological H~{\sc i} signal in delay spectrum based analyses. This is the result of the interplay between foreground emission, particularly diffuse Galactic emission, and the wide-field properties typical of EoR instruments. Our simulations predicted that the delay spectra from the MWA and other experiments should exhibit a characteristic ``pitchfork'' appearance, with local maxima near the horizon delay limits, in addition to at the primary lobe region.  

Careful antenna aperture design can significantly mitigate this contamination. Optimal weighting of contaminated Fourier modes may be required to extract the signal with maximum sensitivity. Thus, detailed knowledge of foreground signatures is key for the design and analysis choices of future instruments such as HERA and SKA.

In Paper~I, we verified the general features of our simulations against MWA observations, but were unable to confirm the {\it pitchfork} prediction due to an insufficient sensitivity in the small amount of data analyzed. Here, we use deeper MWA data to confirm with a high significance the presence of key {\it pitchfork} characteristics of wide-field measurements predicted in the preceding study.  

Section~\ref{sec:wide-field} is an overview of the role of wide-field measurements in the delay spectral domain and the predicted {\it pitchfork} signature. Section~\ref{sec:MWA} describes the analysis of MWA data used to improve the dynamic range of the delay spectra. Section~\ref{sec:results} describes the results and confirms the presence of the predicted wide-field effects. Section~\ref{sec:impact} underscores their impact on aperture design. Section~\ref{sec:summary} summarizes our findings.

\section{Wide-Field Effects in Delay Spectrum}\label{sec:wide-field}

Paper~I describes in detail the effects of wide-field measurements as seen in the delay spectra of interferometer {\it visibilities}. We list the relevant equations and give a brief overview of the wide-field signatures predicted therein. 

The delay spectrum for a baseline vector, $\boldsymbol{b}$, is \citep[][Paper~I]{par12a,par12b,thy13}: 
\begin{align}\label{eqn:delay-spectrum}
  \widetilde{V}_b(\tau) &\equiv \int V_b(f)\,W(f)\,e^{i2\pi f\tau}\,\dif f,
\end{align}
with interferometer visibilities, $V_b(f)$, given by \citep{van34,zer38,tho01}:
\begin{align} 
  V_b(f) &= \iint\limits_\textrm{sky} A(\hat{\boldsymbol{s}},f)\,I(\hat{\boldsymbol{s}},f)\,W_\textrm{i}(f)\,e^{-i2\pi f\frac{\boldsymbol{b}\cdot\hat{\boldsymbol{s}}}{c}}\,\dif\Omega \label{eqn:vis1}\\
         &= \iint\limits_\textrm{sky} \frac{A(\hat{\boldsymbol{s}},f)\,I(\hat{\boldsymbol{s}},f)}{\sqrt{1-l^2-m^2}}\,W_\textrm{i}(f)\,e^{-i2\pi f\frac{\boldsymbol{b}\cdot\hat{\boldsymbol{s}}}{c}}\,\dif l\,\dif m, \label{eqn:vis2}
\end{align}
where, $I(\hat{\boldsymbol{s}},f)$ and $A(\hat{\boldsymbol{s}},f)$ are the sky brightness and antenna's directional power pattern, respectively, as a function of frequency ($f$) and direction on the sky denoted by the unit vector $\hat{\boldsymbol{s}}\equiv (l,m,n)$, $W_\textrm{i}(f)$ denotes instrumental bandpass weights, $W(f)$ is a spectral weighting function that controls the transfer function in the delay transform, $\dif\Omega=(1-l^2-m^2)^{-1/2}\,\dif l\,\dif m$ is the solid angle element to which $\hat{\boldsymbol{s}}$ is the unit normal vector, and $c$ is the speed of light. $\tau=\boldsymbol{b}\cdot\hat{\boldsymbol{s}}/c$ is the geometric delay between antenna pairs measured relative to the zenith and provides a mapping to position on the sky.

The delay power spectrum is defined as \citep[][Paper~I]{par12a}:
\begin{align}\label{eqn:delay-power-spectrum}
  P_\textrm{d}(\boldsymbol{k}_\perp,k_\parallel) &\equiv |\widetilde{V}_b(\tau)|^2\left(\frac{A_\textrm{e}}{\lambda^2\Delta B}\right)\left(\frac{D^2\Delta D}{\Delta B}\right)\left(\frac{\lambda^2}{2k_\textrm{B}}\right)^2,
\end{align}
with
\begin{align}
  \boldsymbol{k}_\perp &\equiv \frac{2\pi(\frac{\boldsymbol{b}}{\lambda})}{D}, \\
  k_\parallel &\equiv \frac{2\pi\tau\,f_{21}H_0\,E(z)}{c\,(1+z)^2}, 
\end{align}
where, $A_\textrm{e}$ is the effective area of the antenna, $\Delta B$ is the bandwidth, $\lambda$ is the wavelength of the band center, $k_\textrm{B}$ is the Boltzmann constant, $f_{21}$ is the rest frequency of the 21~cm radiation of H~{\sc i}, $z$ is the redshift, $D\equiv D(z)$ is the transverse comoving distance, $\Delta D$ is the comoving depth along the line of sight, and $h$, $H_0$ and $E(z)\equiv [\Omega_\textrm{M}(1+z)^3+\Omega_\textrm{k}(1+z)^2+\Omega_\Lambda]^{1/2}$ are standard cosmology terms. In this paper, we use $\Omega_\textrm{M}=0.27$, $\Omega_\Lambda=0.73$, $\Omega_\textrm{K}=1-\Omega_\textrm{M}-\Omega_\Lambda$, and $H_0=100\,$km$\,$s$^{-1}\,$Mpc$^{-1}$. $P_\textrm{d}(\boldsymbol{k}_\perp,k_\parallel)$ is in units of K$^2$(Mpc/$h$)$^3$.

The defining characteristics of the {\it pitchfork} signature are understood as follows. The steep rise in subtended solid angle near the horizon for a fixed delay bin size significantly enhances the integrated emission near the horizon delay limits in wide-field measurements. This is found to be true for diffuse emission even on wide antenna spacings because their foreshortening toward the horizon makes them sensitive to large angular scales that match the inverse of their foreshortened lengths. In the following sections, we present an observational confirmation of this effect.

\section{MWA Observations}\label{sec:MWA}

The MWA instrument configuration, EoR observing strategy, and analysis procedure applied to individual snapshots used in this study are already described in Paper~I and the references therein. In that paper, we analyzed two observations -- an {\it off-zenith} pointing that included significant Galactic plane contributions and a {\it zenith} pointing in which the Galactic plane was significantly absent. The {\it off-zenith} snapshot was useful for demonstrating the mapping between delay spectra and sky locations, establishing primary causes of foreground contamination, and devising a technique for mitigating foreground contamination. {\it Pitchfork} signatures that result from low-level ubiquitous diffuse emission are fainter. Since Galactic plane emission, even from directions far away from the primary field of view, can potentially swamp the fainter {\it pitchfork} signatures, zenith pointings that have a maximum avoidance of the Galactic plane are preferred for this study.

To reduce thermal fluctuations while maintaining coherence, it is essential to average independent data sets obtained over the same region of sky with identical instrument settings. Hence, we select a subset of MWA snapshots pointed at zenith, each with a duration of 112~s, obtained over different nights which are aligned to within 72~s of each other in {\it local sidereal time} (LST) around a mean LST of 0.04~hr. The database contains 14 snapshots satisfying these criteria. Two of these snapshots, which contained amplitude and phase artifacts for a significant duration across different baselines, are excluded from our analysis. The results of this coherent averaging are discussed below.

\section{Results}\label{sec:results}

\begin{figure}[htb]
\centering
\includegraphics[width=\linewidth]{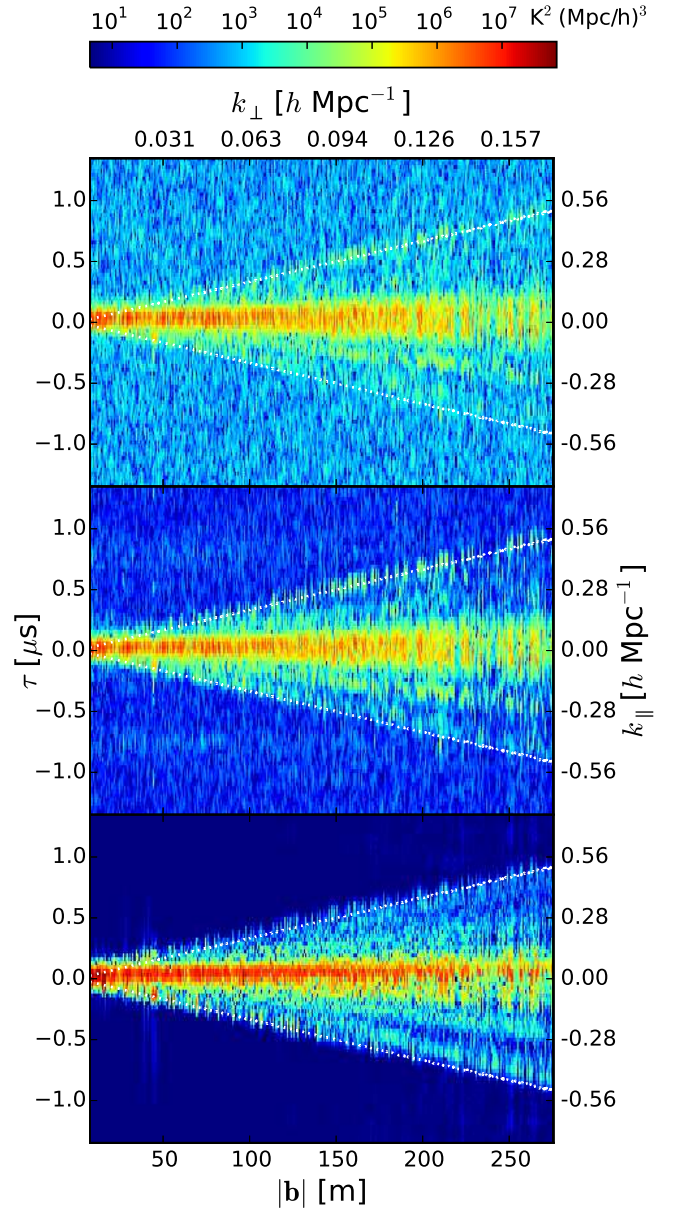}
\caption{Delay power spectra obtained from a single snapshot ({\it top}), by averaging 12 snapshots of LST aligned MWA data ({\it middle}), and from modeling with no thermal noise added ({\it bottom}). The $x$-axis, denoted by $|\boldsymbol{b}|$ (and $k_\perp$), represents angular (and spatial) scales in the plane of the sky while the $y$-axis, shown in $\tau$ and $k_\parallel$, denotes the spatial scales along the line of sight. White dotted lines are the horizon delay limits. Dynamic range in the delay power spectra of MWA data has increased by a factor $\sim 10$ after averaging (middle) relative to that in a single snapshot (top). Power near the horizon limits caused by wide-field effects is prominent. Faint horizontal features at $\tau\simeq\pm 0.78\,\mu$s are visible due to the effective lowering of thermal fluctuations and are the response to periodic coarse band edge flagging of MWA data every 1.28~MHz. \label{fig:delay-spectra}}
\end{figure}

Figure~\ref{fig:delay-spectra} shows the delay spectra obtained from a single snapshot of MWA data (top), averaging LST aligned delay spectra from 12 individual snapshots from MWA observations on different nights (middle) and from modeling with no thermal noise fluctuations shown for reference (bottom). In all panels, the {\it foreground wedge} bounded by horizon limits (white dotted lines) is prominent. The bright horizontal branch of power at $\tau\simeq 0$ corresponds to foreground emission from the main lobe of the antenna power pattern pointed at the zenith. 

In the single snapshot (top), similar to the one used in Paper~I, faint features associated with the {\it pitchfork} signature are visible near the horizon limits. But the high level of thermal fluctuations makes their significance marginal. In contrast, the dynamic range in the averaged data (middle) is a factor $\gtrsim 10$ higher (in the delay power spectrum) relative to that in a single snapshot and is consistent with the improvement expected from averaging 12 independent snapshots. Hence, the foreground power near the horizon limits appears $\gtrsim 10$ times more prominent. Also, faint horizontal features that are not seen in the single snapshot appear at $\tau \simeq \pm 0.78\,\mu$s thus confirming the improvement in sensitivity. We identify these faint features as the response of the MWA coarse band edges flagged periodically every 1.28~MHz. 

In these observations, the Galactic center is just about to set in the west. Its signature on eastward baselines is seen in the modeled delay spectra (bottom panel) as a marginal brightening of the arm near the negative horizon limit for $|\boldsymbol{b}| < 125$~m, consistent with our findings in Paper~I. This spills over into higher delay modes, resulting in the faint ($\lesssim 10^2$~K$^2$~$($Mpc$/h)^3$) vertical stripes at $|\boldsymbol{b}| < 50$~m. The corresponding vertical feature is identified in the averaged data as well.

\begin{figure}[htb]
\centering
\includegraphics[width=\linewidth]{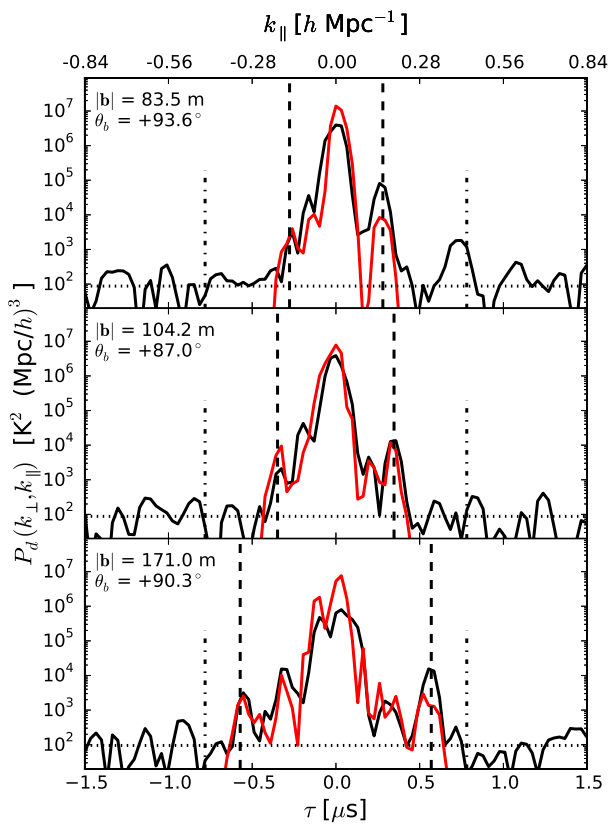}
\caption{Delay power spectra on three antenna spacings oriented northward, obtained by the coherent averaging of 12 snapshots aligned in LST. The averaged data and models are shown in black and red, respectively. The antenna spacings are 83.5~m ({\it top}), 104.2~m ({\it middle}), and 171~m ({\it bottom}). The horizontal dotted line is the {\it rms} of thermal fluctuations. The vertical dashed lines denote the horizon delay limits. The vertical dotted--dashed lines at $\tau = \pm 0.78\,\mu$s correspond to the grating responses of the periodic flagging of bandpass at intervals of 1.28~MHz. The peaks close to the horizon delay limits are distinctly visible at $\sim 10$--1000~$\sigma$ levels. Differences between the model and the data are primarily attributed to uncertainties in the foreground model and the MWA tile power pattern. \label{fig:3-baseline-comparison-delay-spectra}}
\end{figure}

In order to show that low-level ubiquitous diffuse emission is a significant contributor to the {\it pitchfork} signature, contribution from any strong emission from near the Galactic center needs to be minimized. This is best illustrated with northward antenna spacings that map any residual emission from this region to $\tau\simeq 0$ and thus reduce the impact on higher delay modes (Paper~I). Figure~\ref{fig:3-baseline-comparison-delay-spectra} shows the averaged delay power spectra on three selected baseline vectors oriented northward. Data and noiseless models are shown in black and red, respectively. The horizontal dotted black line denotes the {\it rms} of thermal fluctuations estimated from data. The vertical dashed line denotes horizon delay limits, and the vertical dotted--dashed lines denote delays at which the responses to coarse band edge flagging are expected.

We focus on the prominent peaks in data near the horizon limits. Typically, the power near the negative horizon limit is seen with a signal-to-noise ratio $\sim$~10--100, while that around the positive horizon limit is $\sim$~100--1000. 

There is a remarkable agreement in broad morphology between the data and the model. However, some differences in the amplitude scales are noted. For instance, the emission near the positive horizon limit is higher in the data than in the model in both figures. We attribute such differences to uncertainties in the foreground model, the MWA tile power pattern, thermal fluctuations, and other uncertainties noted in Paper~I which limit a more thorough quantitative comparison between the model and the data. In fact, this lays further emphasis on the need for the following -- extending the footprint of surveys matching the frequency and angular resolution of observations such as the MWA Commissioning Survey \citep[MWACS;][]{hur14} and the Galactic and Extragalactic MWA Survey \citep[GLEAM;][]{way15} to cover the entire hemisphere, and a detailed characterization of the power pattern over the entire hemisphere comprehensively extending studies such as those in \citet{neb15}.

We note that reducing uncertainties will only change the relative strength of the {\it pitchfork} signature in our model. However, the effects giving rise to this signature are generic to all wide-field measurements. Thus, the extremely high significance detection of foreground emission near the horizon limits is a robust confirmation of the predicted effects of wide-field measurements.

\section{Impact on Instrument Design}\label{sec:impact}

The delay spectrum maps the geometric delays to positions of foreground objects on the sky. Thus the directional power pattern of the antenna has a direct impact on the delay spectrum. In fact, from equations~\ref{eqn:delay-spectrum}, \ref{eqn:vis1}, and \ref{eqn:delay-power-spectrum}, the delay power spectrum scales as the square of the directional power pattern of the antenna. Since the contamination in the {\it EoR window} is strongly dependent on sources of emission close to the horizon, the design of antenna apertures that suppress their sensitivity toward the horizon is important. 

We approached this issue in Paper~I via simulations of different antenna apertures -- a dipole, a $4\times 4$ phased array, and a dish. Among these, a dish provides the best quality, as evaluated from a foreground contamination viewpoint. With new instruments such as HERA and SKA on the horizon, we extend this discussion to further emphasize the need to comprehensively study the effect of their antenna apertures on foreground contamination. 

HERA will deploy 14~m dishes, with 331 of them in a closely packed hexagonal array and another 21 as outliers. With the advantage of enormous redundancy and unprecedented sensitivity, it will address the following key questions -- what objects first lit up the Universe and reionized the neutral IGM? Over what redshift range did this occur? And how did the process proceed, leading to the large-scale galaxy structure seen today? The fixed dishes will observe the sky drifting overhead. Based on simulations of uniformly illuminated circular disks, the response near the horizon is $\lesssim -40$~dB, which is over 20~dB lower than that of the MWA. Tapering the reflectivity of the dish is being studied to further reduce this response. The responses of the actual dishes will be tested shortly. 

With the SKA at low frequencies, deploying ``stations'' of $\sim$~35~m diameter that act as aperture arrays consisting of 256 pseudo-randomly placed vertical log-periodic dipole antennas is being considered.\footnote{Documents at \url{https://www.skatelescope.org/key-documents/} on SKA re-baselining}$^,$\footnote{SKA memo ``Station Response and Imaging Performance of LFAA: 100 MHz to 600 MHz'' by Razavi-Ghods et al.} The voltage beam of the station will be obtained by a phased addition of the dipole responses. Since the power pattern of a baseline consisting of two stations will be effectively be a product of two voltage beams, it will homogenize the fluctuations in the power patterns of individual stations. Even with simple models of such phased array stations, we estimate the typical horizon response of these beams to be $-30$~dB to $-40$~dB compared to the zenith, yielding a response that is at least 10~dB better than the MWA. A targeted optimization of the antenna layouts of stations is under active study and could yield even better responses. 

From these simple antenna aperture models, HERA and SKA should mitigate leakage from the {\it pitchfork} into the {\it EoR window} by $\sim 40$~dB relative to the MWA when expressed in units of power spectrum. This will significantly increase the likelihood of detecting the cosmological H~{\sc i} power spectrum in sensitive three-dimensional $k$-modes adjacent to the {\it pitchfork}. The precise beam responses and the resulting foreground leakage, especially from near the horizon, will prove to be critical and timely inputs to the actual aperture designs of HERA dishes and SKA stations, which are currently nearing their final stages.

\section{Summary}\label{sec:summary}

Using deeper MWA data, we have confirmed with a high significance the earlier prediction of a characteristic {\it pitchfork} morphology where wide-field EoR measurements suffer from significant foreground contamination from near the horizon. This has important implications for the instrument design and data analysis of future instruments such as HERA and SKA. 

Careful aperture designs that suppress the response near the horizon, and hence also the leakage from the {\it pitchfork}, will significantly avoid contamination in $k$-modes considered critical for cosmological signal detection. Precise modeling is thus required to gain a complete understanding of the characteristics of the cosmological signal and the foregrounds. 

Foreground and instrument models serve as inputs for power spectrum estimation techniques \citep[see, e.g.,][]{liu14a,liu14b,dil15}. Input models that ignore wide-field effects will provide over-valued weights in and around the {\it pitchfork} modes, and hence result in sub-optimal results. Thus, confirmation of the {\it pitchfork} effect has important implications for optimum power spectrum estimation. 

In Paper~I, we proposed a selective flagging of data on different baselines that can potentially mitigate foreground contamination by two orders of magnitude. Following the confirmation presented here, efforts are underway to incorporate this proposed foreground mitigation technique into the MWA data analysis.

For future work, we plan to extend our analysis to HERA. Based on Paper~I, a dish will have a much desirable Fourier response from a foreground contamination viewpoint. One of our objectives is to forecast the per-baseline foreground contamination as a function of LST in order to tune the HERA observing strategy and data analysis to maximize sensitivity to the EoR signal.

\acknowledgments

This work was supported by the U. S. National Science Foundation (NSF) through award AST-1109257. D.C.J. is supported by an NSF Astronomy and Astrophysics Postdoctoral Fellowship under award AST-1401708. J.C.P. is supported by an NSF Astronomy and Astrophysics Fellowship under award AST-1302774. This work makes use of the Murchison Radio-astronomy Observatory, operated by CSIRO. We acknowledge the Wajarri Yamatji people as the traditional owners of the observatory site. Support for the MWA comes from the NSF (awards: AST-0457585, PHY-0835713, CAREER-0847753, and AST-0908884), the Australian Research Council (LIEF grants LE0775621 and LE0882938), the U.S. Air Force Office of Scientific Research (grant FA9550-0510247), and the Centre for All-sky Astrophysics (an Australian Research Council Centre of Excellence funded by grant CE110001020). Support is also provided by the Smithsonian Astrophysical Observatory, the MIT School of Science, the Raman Research Institute, the Australian National University, and the Victoria University of Wellington (via grant MED-E1799 from the New Zealand Ministry of Economic Development and an IBM Shared University Research Grant). The Australian Federal government provides additional support via the Commonwealth Scientific and Industrial Research Organisation (CSIRO), National Collaborative Research Infrastructure Strategy, Education Investment Fund, and the Australia India Strategic Research Fund, and Astronomy Australia Limited, under contract to Curtin University. We acknowledge the iVEC Petabyte Data Store, the Initiative in Innovative Computing and the CUDA Center for Excellence sponsored by NVIDIA at Harvard University, and the International Centre for Radio Astronomy Research (ICRAR), a Joint Venture of Curtin University and The University of Western Australia, funded by the Western Australian State government.  

\bibliographystyle{apj}

\begin{thebibliography}{}
\expandafter\ifx\csname natexlab\endcsname\relax\def\natexlab#1{#1}\fi

\bibitem[{{Ali} {et~al.}(2008){Ali}, {Bharadwaj}, \& {Chengalur}}]{ali08}
{Ali}, S.~S., {Bharadwaj}, S., \& {Chengalur}, J.~N. 2008, \mnras, 385, 2166

\bibitem[{{Beardsley} {et~al.}(2013){Beardsley}, {Hazelton}, {Morales},
  {Arcus}, {Barnes}, {Bernardi}, {Bowman}, {Briggs}, {Bunton}, {Cappallo},
  {Corey}, {Deshpande}, {deSouza}, {Emrich}, {Gaensler}, {Goeke}, {Greenhill},
  {Herne}, {Hewitt}, {Johnston-Hollitt}, {Kaplan}, {Kasper}, {Kincaid},
  {Koenig}, {Kratzenberg}, {Lonsdale}, {Lynch}, {McWhirter}, {Mitchell},
  {Morgan}, {Oberoi}, {Ord}, {Pathikulangara}, {Prabu}, {Remillard}, {Rogers},
  {Roshi}, {Salah}, {Sault}, {Udaya}, {Srivani}, {Stevens}, {Subrahmanyan},
  {Tingay}, {Wayth}, {Waterson}, {Webster}, {Whitney}, {Williams}, {Williams},
  \& {Wyithe}}]{bea13}
{Beardsley}, A.~P., {Hazelton}, B.~J., {Morales}, M.~F., {et~al.} 2013, \mnras,
  429, L5

\bibitem[{{Bernardi} {et~al.}(2009){Bernardi}, {de Bruyn}, {Brentjens},
  {Ciardi}, {Harker}, {Jeli{\'c}}, {Koopmans}, {Labropoulos}, {Offringa},
  {Pandey}, {Schaye}, {Thomas}, {Yatawatta}, \& {Zaroubi}}]{ber09}
{Bernardi}, G., {de Bruyn}, A.~G., {Brentjens}, M.~A., {et~al.} 2009, \aap,
  500, 965

\bibitem[{{Bernardi} {et~al.}(2010){Bernardi}, {de Bruyn}, {Harker},
  {Brentjens}, {Ciardi}, {Jeli{\'c}}, {Koopmans}, {Labropoulos}, {Offringa},
  {Pandey}, {Schaye}, {Thomas}, {Yatawatta}, \& {Zaroubi}}]{ber10}
{Bernardi}, G., {de Bruyn}, A.~G., {Harker}, G., {et~al.} 2010, \aap, 522, A67

\bibitem[{{Bowman} {et~al.}(2006){Bowman}, {Morales}, \& {Hewitt}}]{bow06}
{Bowman}, J.~D., {Morales}, M.~F., \& {Hewitt}, J.~N. 2006, \apj, 638, 20

\bibitem[{{Bowman} {et~al.}(2009){Bowman}, {Morales}, \& {Hewitt}}]{bow09}
---. 2009, \apj, 695, 183

\bibitem[{{Bowman} {et~al.}(2013){Bowman}, {Cairns}, {Kaplan}, {Murphy},
  {Oberoi}, {Staveley-Smith}, {Arcus}, {Barnes}, {Bernardi}, {Briggs}, {Brown},
  {Bunton}, {Burgasser}, {Cappallo}, {Chatterjee}, {Corey}, {Coster},
  {Deshpande}, {deSouza}, {Emrich}, {Erickson}, {Goeke}, {Gaensler},
  {Greenhill}, {Harvey-Smith}, {Hazelton}, {Herne}, {Hewitt},
  {Johnston-Hollitt}, {Kasper}, {Kincaid}, {Koenig}, {Kratzenberg}, {Lonsdale},
  {Lynch}, {Matthews}, {McWhirter}, {Mitchell}, {Morales}, {Morgan}, {Ord},
  {Pathikulangara}, {Prabu}, {Remillard}, {Robishaw}, {Rogers}, {Roshi},
  {Salah}, {Sault}, {Shankar}, {Srivani}, {Stevens}, {Subrahmanyan}, {Tingay},
  {Wayth}, {Waterson}, {Webster}, {Whitney}, {Williams}, {Williams}, \&
  {Wyithe}}]{bow13}
{Bowman}, J.~D., {Cairns}, I., {Kaplan}, D.~L., {et~al.} 2013, \pasa, 30, 31

\bibitem[{{Datta} {et~al.}(2010){Datta}, {Bowman}, \& {Carilli}}]{dat10}
{Datta}, A., {Bowman}, J.~D., \& {Carilli}, C.~L. 2010, \apj, 724, 526

\bibitem[{{Di Matteo} {et~al.}(2004){Di Matteo}, {Ciardi}, \&
  {Miniati}}]{dim04}
{Di Matteo}, T., {Ciardi}, B., \& {Miniati}, F. 2004, \mnras, 355, 1053

\bibitem[{{Di Matteo} {et~al.}(2002){Di Matteo}, {Perna}, {Abel}, \&
  {Rees}}]{dim02}
{Di Matteo}, T., {Perna}, R., {Abel}, T., \& {Rees}, M.~J. 2002, \apj, 564, 576

\bibitem[{Dillon {et~al.}(2013)Dillon, Liu, \& Tegmark}]{dil13}
Dillon, J.~S., Liu, A., \& Tegmark, M. 2013, Phys. Rev. D, 87, 043005

\bibitem[{{Dillon} {et~al.}(2014){Dillon}, {Liu}, {Williams}, {Hewitt},
  {Tegmark}, {Morgan}, {Levine}, {Morales}, {Tingay}, {Bernardi}, {Bowman},
  {Briggs}, {Cappallo}, {Emrich}, {Mitchell}, {Oberoi}, {Prabu}, {Wayth}, \&
  {Webster}}]{dil14}
{Dillon}, J.~S., {Liu}, A., {Williams}, C.~L., {et~al.} 2014, \prd, 89, 023002

\bibitem[{{Dillon} {et~al.}(2015){Dillon}, {Neben}, {Hewitt}, {Tegmark},
  {Barry}, {Beardsley}, {Bowman}, {Briggs}, {Carroll}, {de Oliveira-Costa},
  {Ewall-Wice}, {Feng}, {Greenhill}, {Hazelton}, {Hernquist}, {Hurley-Walker},
  {Jacobs}, {Kim}, {Kittiwisit}, {Lenc}, {Line}, {Loeb}, {McKinley},
  {Mitchell}, {Morales}, {Offringa}, {Paul}, {Pindor}, {Pober}, {Procopio},
  {Riding}, {Sethi}, {Udaya Shankar}, {Subrahmanyan}, {Sullivan},
  {Thyagarajan}, {Tingay}, {Trott}, {Wayth}, {Webster}, {Wyithe}, {Bernardi},
  {Cappallo}, {Deshpande}, {Johnston-Hollitt}, {Kaplan}, {Lonsdale},
  {McWhirter}, {Morgan}, {Oberoi}, {Ord}, {Prabu}, {Srivani}, {Williams}, \&
  {Williams}}]{dil15}
{Dillon}, J.~S., {Neben}, A.~R., {Hewitt}, J.~N., {et~al.} 2015, ArXiv
  e-prints, arXiv:1506.01026

\bibitem[{{Furlanetto} \& {Briggs}(2004)}]{fur04}
{Furlanetto}, S.~R., \& {Briggs}, F.~H. 2004, \nar, 48, 1039

\bibitem[{{Furlanetto} {et~al.}(2006){Furlanetto}, {Oh}, \& {Briggs}}]{fur06}
{Furlanetto}, S.~R., {Oh}, S.~P., \& {Briggs}, F.~H. 2006, \physrep, 433, 181

\bibitem[{{Ghosh} {et~al.}(2012){Ghosh}, {Prasad}, {Bharadwaj}, {Ali}, \&
  {Chengalur}}]{gho12}
{Ghosh}, A., {Prasad}, J., {Bharadwaj}, S., {Ali}, S.~S., \& {Chengalur}, J.~N.
  2012, \mnras, 426, 3295

\bibitem[{{Gleser} {et~al.}(2008){Gleser}, {Nusser}, \& {Benson}}]{gle08}
{Gleser}, L., {Nusser}, A., \& {Benson}, A.~J. 2008, \mnras, 391, 383

\bibitem[{{Hurley-Walker} {et~al.}(2014){Hurley-Walker}, {Morgan}, {Wayth},
  {Hancock}, {Bell}, {Bernardi}, {Bhat}, {Briggs}, {Deshpande}, {Ewall-Wice},
  {Feng}, {Hazelton}, {Hindson}, {Jacobs}, {Kaplan}, {Kudryavtseva}, {Lenc},
  {McKinley}, {Mitchell}, {Pindor}, {Procopio}, {Oberoi}, {Offringa}, {Ord},
  {Riding}, {Bowman}, {Cappallo}, {Corey}, {Emrich}, {Gaensler}, {Goeke},
  {Greenhill}, {Hewitt}, {Johnston-Hollitt}, {Kasper}, {Kratzenberg},
  {Lonsdale}, {Lynch}, {McWhirter}, {Morales}, {Morgan}, {Prabu}, {Rogers},
  {Roshi}, {Shankar}, {Srivani}, {Subrahmanyan}, {Tingay}, {Waterson},
  {Webster}, {Whitney}, {Williams}, \& {Williams}}]{hur14}
{Hurley-Walker}, N., {Morgan}, J., {Wayth}, R.~B., {et~al.} 2014, \pasa, 31, 45

\bibitem[{{Iliev} {et~al.}(2002){Iliev}, {Shapiro}, {Ferrara}, \&
  {Martel}}]{ili02}
{Iliev}, I.~T., {Shapiro}, P.~R., {Ferrara}, A., \& {Martel}, H. 2002, \apjl,
  572, L123

\bibitem[{{Liu} {et~al.}(2014{\natexlab{a}}){Liu}, {Parsons}, \&
  {Trott}}]{liu14a}
{Liu}, A., {Parsons}, A.~R., \& {Trott}, C.~M. 2014{\natexlab{a}}, \prd, 90,
  023018

\bibitem[{{Liu} {et~al.}(2014{\natexlab{b}}){Liu}, {Parsons}, \&
  {Trott}}]{liu14b}
---. 2014{\natexlab{b}}, \prd, 90, 023019

\bibitem[{{Liu} \& {Tegmark}(2011)}]{liu11}
{Liu}, A., \& {Tegmark}, M. 2011, \prd, 83, 103006

\bibitem[{{Liu} {et~al.}(2009){Liu}, {Tegmark}, {Bowman}, {Hewitt}, \&
  {Zaldarriaga}}]{liu09}
{Liu}, A., {Tegmark}, M., {Bowman}, J., {Hewitt}, J., \& {Zaldarriaga}, M.
  2009, \mnras, 398, 401

\bibitem[{{Lonsdale} {et~al.}(2009){Lonsdale}, {Cappallo}, {Morales}, {Briggs},
  {Benkevitch}, {Bowman}, {Bunton}, {Burns}, {Corey}, {Desouza}, {Doeleman},
  {Derome}, {Deshpande}, {Gopala}, {Greenhill}, {Herne}, {Hewitt}, {Kamini},
  {Kasper}, {Kincaid}, {Kocz}, {Kowald}, {Kratzenberg}, {Kumar}, {Lynch},
  {Madhavi}, {Matejek}, {Mitchell}, {Morgan}, {Oberoi}, {Ord},
  {Pathikulangara}, {Prabu}, {Rogers}, {Roshi}, {Salah}, {Sault}, {Shankar},
  {Srivani}, {Stevens}, {Tingay}, {Vaccarella}, {Waterson}, {Wayth}, {Webster},
  {Whitney}, {Williams}, \& {Williams}}]{lon09}
{Lonsdale}, C.~J., {Cappallo}, R.~J., {Morales}, M.~F., {et~al.} 2009, IEEE
  Proceedings, 97, 1497

\bibitem[{{Madau} {et~al.}(1997){Madau}, {Meiksin}, \& {Rees}}]{mad97}
{Madau}, P., {Meiksin}, A., \& {Rees}, M.~J. 1997, \apj, 475, 429

\bibitem[{{McQuinn} {et~al.}(2006){McQuinn}, {Zahn}, {Zaldarriaga},
  {Hernquist}, \& {Furlanetto}}]{mcq06}
{McQuinn}, M., {Zahn}, O., {Zaldarriaga}, M., {Hernquist}, L., \& {Furlanetto},
  S.~R. 2006, \apj, 653, 815

\bibitem[{{Morales} {et~al.}(2006){Morales}, {Bowman}, \& {Hewitt}}]{mor06}
{Morales}, M.~F., {Bowman}, J.~D., \& {Hewitt}, J.~N. 2006, \apj, 648, 767

\bibitem[{{Morales} {et~al.}(2012){Morales}, {Hazelton}, {Sullivan}, \&
  {Beardsley}}]{mor12}
{Morales}, M.~F., {Hazelton}, B., {Sullivan}, I., \& {Beardsley}, A. 2012,
  \apj, 752, 137

\bibitem[{{Morales} \& {Hewitt}(2004)}]{mor04}
{Morales}, M.~F., \& {Hewitt}, J. 2004, \apj, 615, 7

\bibitem[{{Neben} {et~al.}(2015){Neben}, {Bradley}, {Hewitt}, {Bernardi},
  {Bowman}, {Briggs}, {Cappallo}, {Deshpande}, {Goeke}, {Greenhill},
  {Hazelton}, {Johnston-Hollitt}, {Kaplan}, {Lonsdale}, {McWhirter},
  {Mitchell}, {Morales}, {Morgan}, {Oberoi}, {Ord}, {Prabu}, {Udaya Shankar},
  {Srivani}, {Subrahmanyan}, {Tingay}, {Wayth}, {Webster}, {Williams}, \&
  {Williams}}]{neb15}
{Neben}, A.~R., {Bradley}, R.~F., {Hewitt}, J.~N., {et~al.} 2015, ArXiv
  e-prints, arXiv:1505.07114

\bibitem[{{Parsons} {et~al.}(2012{\natexlab{a}}){Parsons}, {Pober}, {McQuinn},
  {Jacobs}, \& {Aguirre}}]{par12a}
{Parsons}, A., {Pober}, J., {McQuinn}, M., {Jacobs}, D., \& {Aguirre}, J.
  2012{\natexlab{a}}, \apj, 753, 81

\bibitem[{{Parsons} {et~al.}(2012{\natexlab{b}}){Parsons}, {Pober}, {Aguirre},
  {Carilli}, {Jacobs}, \& {Moore}}]{par12b}
{Parsons}, A.~R., {Pober}, J.~C., {Aguirre}, J.~E., {et~al.}
  2012{\natexlab{b}}, \apj, 756, 165

\bibitem[{{Parsons} {et~al.}(2010){Parsons}, {Backer}, {Foster}, {Wright},
  {Bradley}, {Gugliucci}, {Parashare}, {Benoit}, {Aguirre}, {Jacobs},
  {Carilli}, {Herne}, {Lynch}, {Manley}, \& {Werthimer}}]{par10}
{Parsons}, A.~R., {Backer}, D.~C., {Foster}, G.~S., {et~al.} 2010, \aj, 139,
  1468

\bibitem[{{Pober} {et~al.}(2013){Pober}, {Parsons}, {Aguirre}, {Ali},
  {Bradley}, {Carilli}, {DeBoer}, {Dexter}, {Gugliucci}, {Jacobs}, {Klima},
  {MacMahon}, {Manley}, {Moore}, {Stefan}, \& {Walbrugh}}]{pob13}
{Pober}, J.~C., {Parsons}, A.~R., {Aguirre}, J.~E., {et~al.} 2013, \apjl, 768,
  L36

\bibitem[{{Pober} {et~al.}(2014){Pober}, {Liu}, {Dillon}, {Aguirre}, {Bowman},
  {Bradley}, {Carilli}, {DeBoer}, {Hewitt}, {Jacobs}, {McQuinn}, {Morales},
  {Parsons}, {Tegmark}, \& {Werthimer}}]{pob14}
{Pober}, J.~C., {Liu}, A., {Dillon}, J.~S., {et~al.} 2014, \apj, 782, 66

\bibitem[{{Santos} {et~al.}(2005){Santos}, {Cooray}, \& {Knox}}]{san05}
{Santos}, M.~G., {Cooray}, A., \& {Knox}, L. 2005, \apj, 625, 575

\bibitem[{{Scott} \& {Rees}(1990)}]{sco90}
{Scott}, D., \& {Rees}, M.~J. 1990, \mnras, 247, 510

\bibitem[{{Sunyaev} \& {Zeldovich}(1972)}]{sun72}
{Sunyaev}, R.~A., \& {Zeldovich}, Y.~B. 1972, \aap, 20, 189

\bibitem[{{Thompson} {et~al.}(2001){Thompson}, {Moran}, \& {Swenson}}]{tho01}
{Thompson}, A.~R., {Moran}, J.~M., \& {Swenson}, Jr., G.~W. 2001,
  {Interferometry and Synthesis in Radio Astronomy, 2nd Edition} (Wiley)

\bibitem[{{Thyagarajan} {et~al.}(2013){Thyagarajan}, {Udaya Shankar},
  {Subrahmanyan}, {Arcus}, {Bernardi}, {Bowman}, {Briggs}, {Bunton},
  {Cappallo}, {Corey}, {deSouza}, {Emrich}, {Gaensler}, {Goeke}, {Greenhill},
  {Hazelton}, {Herne}, {Hewitt}, {Johnston-Hollitt}, {Kaplan}, {Kasper},
  {Kincaid}, {Koenig}, {Kratzenberg}, {Lonsdale}, {Lynch}, {McWhirter},
  {Mitchell}, {Morales}, {Morgan}, {Oberoi}, {Ord}, {Pathikulangara},
  {Remillard}, {Rogers}, {Anish Roshi}, {Salah}, {Sault}, {Srivani}, {Stevens},
  {Thiagaraj}, {Tingay}, {Wayth}, {Waterson}, {Webster}, {Whitney}, {Williams},
  {Williams}, \& {Wyithe}}]{thy13}
{Thyagarajan}, N., {Udaya Shankar}, N., {Subrahmanyan}, R., {et~al.} 2013,
  \apj, 776, 6

\bibitem[{{Thyagarajan} {et~al.}(2015){Thyagarajan}, {Jacobs}, {Bowman},
  {Barry}, {Beardsley}, {Bernardi}, {Briggs}, {Cappallo}, {Carroll}, {Corey},
  {de Oliveira-Costa}, {Dillon}, {Emrich}, {Ewall-Wice}, {Feng}, {Goeke},
  {Greenhill}, {Hazelton}, {Hewitt}, {Hurley-Walker}, {Johnston-Hollitt},
  {Kaplan}, {Kasper}, {Kim}, {Kittiwisit}, {Kratzenberg}, {Lenc}, {Line},
  {Loeb}, {Lonsdale}, {Lynch}, {McKinley}, {McWhirter}, {Mitchell}, {Morales},
  {Morgan}, {Neben}, {Oberoi}, {Offringa}, {Ord}, {Paul}, {Pindor}, {Pober},
  {Prabu}, {Procopio}, {Riding}, {Rogers}, {Roshi}, {Udaya Shankar}, {Sethi},
  {Srivani}, {Subrahmanyan}, {Sullivan}, {Tegmark}, {Tingay}, {Trott},
  {Waterson}, {Wayth}, {Webster}, {Whitney}, {Williams}, {Williams}, {Wu}, \&
  {Wyithe}}]{thy15}
{Thyagarajan}, N., {Jacobs}, D.~C., {Bowman}, J.~D., {et~al.} 2015, \apj, 804,
  14

\bibitem[{{Tingay} {et~al.}(2013){Tingay}, {Goeke}, {Bowman}, {Emrich}, {Ord},
  {Mitchell}, {Morales}, {Booler}, {Crosse}, {Wayth}, {Lonsdale}, {Tremblay},
  {Pallot}, {Colegate}, {Wicenec}, {Kudryavtseva}, {Arcus}, {Barnes},
  {Bernardi}, {Briggs}, {Burns}, {Bunton}, {Cappallo}, {Corey}, {Deshpande},
  {Desouza}, {Gaensler}, {Greenhill}, {Hall}, {Hazelton}, {Herne}, {Hewitt},
  {Johnston-Hollitt}, {Kaplan}, {Kasper}, {Kincaid}, {Koenig}, {Kratzenberg},
  {Lynch}, {Mckinley}, {Mcwhirter}, {Morgan}, {Oberoi}, {Pathikulangara},
  {Prabu}, {Remillard}, {Rogers}, {Roshi}, {Salah}, {Sault}, {Udaya-Shankar},
  {Schlagenhaufer}, {Srivani}, {Stevens}, {Subrahmanyan}, {Waterson},
  {Webster}, {Whitney}, {Williams}, {Williams}, \& {Wyithe}}]{tin13}
{Tingay}, S.~J., {Goeke}, R., {Bowman}, J.~D., {et~al.} 2013, \pasa, 30, 7

\bibitem[{{Tozzi} {et~al.}(2000){Tozzi}, {Madau}, {Meiksin}, \& {Rees}}]{toz00}
{Tozzi}, P., {Madau}, P., {Meiksin}, A., \& {Rees}, M.~J. 2000, \apj, 528, 597

\bibitem[{{Trott} {et~al.}(2012){Trott}, {Wayth}, \& {Tingay}}]{tro12}
{Trott}, C.~M., {Wayth}, R.~B., \& {Tingay}, S.~J. 2012, \apj, 757, 101

\bibitem[{{van Cittert}(1934)}]{van34}
{van Cittert}, P.~H. 1934, Physica, 1, 201

\bibitem[{{van Haarlem} {et~al.}(2013){van Haarlem}, {Wise}, {Gunst}, {Heald},
  {McKean}, {Hessels}, {de Bruyn}, {Nijboer}, {Swinbank}, {Fallows},
  {Brentjens}, {Nelles}, {Beck}, {Falcke}, {Fender}, {H{\"o}randel},
  {Koopmans}, {Mann}, {Miley}, {R{\"o}ttgering}, {Stappers}, {Wijers},
  {Zaroubi}, {van den Akker}, {Alexov}, {Anderson}, {Anderson}, {van Ardenne},
  {Arts}, {Asgekar}, {Avruch}, {Batejat}, {B{\"a}hren}, {Bell}, {Bell}, {van
  Bemmel}, {Bennema}, {Bentum}, {Bernardi}, {Best}, {B{\^i}rzan}, {Bonafede},
  {Boonstra}, {Braun}, {Bregman}, {Breitling}, {van de Brink}, {Broderick},
  {Broekema}, {Brouw}, {Br{\"u}ggen}, {Butcher}, {van Cappellen}, {Ciardi},
  {Coenen}, {Conway}, {Coolen}, {Corstanje}, {Damstra}, {Davies}, {Deller},
  {Dettmar}, {van Diepen}, {Dijkstra}, {Donker}, {Doorduin}, {Dromer}, {Drost},
  {van Duin}, {Eisl{\"o}ffel}, {van Enst}, {Ferrari}, {Frieswijk}, {Gankema},
  {Garrett}, {de Gasparin}, {Gerbers}, {de Geus}, {Grie{\ss}meier}, {Grit},
  {Gruppen}, {Hamaker}, {Hassall}, {Hoeft}, {Holties}, {Horneffer}, {van der
  Horst}, {van Houwelingen}, {Huijgen}, {Iacobelli}, {Intema}, {Jackson},
  {Jelic}, {de Jong}, {Kant}, {Karastergiou}, {Koers}, {Kollen}, {Kondratiev},
  {Kooistra}, {Koopman}, {Koster}, {Kuniyoshi}, {Kramer}, {Kuper},
  {Lambropoulos}, {Law}, {van Leeuwen}, {Lemaitre}, {Loose}, {Maat}, {Macario},
  {Markoff}, {Masters}, {McFadden}, {McKay-Bukowski}, {Meijering}, {Meulman},
  {Mevius}, {Millenaar}, {Miller-Jones}, {Mohan}, {Mol}, {Morawietz},
  {Morganti}, {Mulcahy}, {Mulder}, {Munk}, {Nieuwenhuis}, {van Nieuwpoort},
  {Noordam}, {Norden}, {Noutsos}, {Offringa}, {Olofsson}, {Omar}, {Orr{\'u}},
  {Overeem}, {Paas}, {Pandey-Pommier}, {Pandey}, {Pizzo}, {Polatidis},
  {Rafferty}, {Rawlings}, {Reich}, {de Reijer}, {Reitsma}, {Renting},
  {Riemers}, {Rol}, {Romein}, {Roosjen}, {Ruiter}, {Scaife}, {van der Schaaf},
  {Scheers}, {Schellart}, {Schoenmakers}, {Schoonderbeek}, {Serylak},
  {Shulevski}, {Sluman}, {Smirnov}, {Sobey}, {Spreeuw}, {Steinmetz}, {Sterks},
  {Stiepel}, {Stuurwold}, {Tagger}, {Tang}, {Tasse}, {Thomas}, {Thoudam},
  {Toribio}, {van der Tol}, {Usov}, {van Veelen}, {van der Veen}, {ter Veen},
  {Verbiest}, {Vermeulen}, {Vermaas}, {Vocks}, {Vogt}, {de Vos}, {van der Wal},
  {van Weeren}, {Weggemans}, {Weltevrede}, {White}, {Wijnholds}, {Wilhelmsson},
  {Wucknitz}, {Yatawatta}, {Zarka}, {Zensus}, \& {van Zwieten}}]{van13}
{van Haarlem}, M.~P., {Wise}, M.~W., {Gunst}, A.~W., {et~al.} 2013, \aap, 556,
  A2

\bibitem[{{Wang} {et~al.}(2006){Wang}, {Tegmark}, {Santos}, \& {Knox}}]{wan06}
{Wang}, X., {Tegmark}, M., {Santos}, M.~G., \& {Knox}, L. 2006, \apj, 650, 529

\bibitem[{{Wayth} {et~al.}(2015){Wayth}, {Lenc}, {Bell}, {Callingham},
  {Dwarakanath}, {Franzen}, {For}, {Gaensler}, {Hancock}, {Hindson},
  {Hurley-Walker}, {Jackson}, {Johnston-Hollitt}, {Kapinska}, {McKinley},
  {Morgan}, {Offringa}, {Procopio}, {Staveley-Smith}, {Wu}, {Zheng}, {Trott},
  {Bernardi}, {Bowman}, {Briggs}, {Cappallo}, {Corey}, {Deshpande}, {Emrich},
  {Goeke}, {Greenhill}, {Hazelton}, {Kaplan}, {Kasper}, {Kratzenberg},
  {Lonsdale}, {Lynch}, {McWhirter}, {Mitchell}, {Morales}, {Morgan}, {Oberoi},
  {Ord}, {Prabu}, {Rogers}, {Roshi}, {Udaya Shankar}, {Srivani},
  {Subrahmanyan}, {Tingay}, {Waterson}, {Webster}, {Whitney}, {Williams}, \&
  {Williams}}]{way15}
{Wayth}, R.~B., {Lenc}, E., {Bell}, M.~E., {et~al.} 2015, ArXiv e-prints,
  arXiv:1505.06041

\bibitem[{{Zaldarriaga} {et~al.}(2004){Zaldarriaga}, {Furlanetto}, \&
  {Hernquist}}]{zal04}
{Zaldarriaga}, M., {Furlanetto}, S.~R., \& {Hernquist}, L. 2004, \apj, 608, 622

\bibitem[{{Zernike}(1938)}]{zer38}
{Zernike}, F. 1938, Physica, 5, 785

\end{thebibliography}

\end{document}